%Paper: hep-th/9207056
%From: LAYNAM@vtcc1.cc.vt.edu (LNC)
%Date: Wed, 15 Jul 1992 23:35:26 -0400 (EDT)
%Date (revised): Thu, 30 Jul 1992 3:22:25 -0400 (EDT)

%%%
\input harvmac
%\draftmode
\def\As{{\rm Ashtekar}}
\def\ts{\tilde{\sigma}}
\def\ut#1{\rlap{\lower1ex\hbox{$\sim$}}#1{}}
\def\al{\alpha}

\def\eps{\epsilon}
\def\cM{{\cal M}}
\def\cD{{\cal D}}
\def\cW{{\cal W}}
\def\cMM{{\cal M} ^{4}}
\def\ha{{1\over 2}}
\def\w{\wedge}
\def\Tr{\hbox{tr}\,}

\def\Ttitle#1#2#3{\nopagenumbers\abstractfont\hsize=\hstitle\rightline{#1}%
\nopagenumbers\abstractfont\hsize=\hstitle\rightline{#2}%
\vskip 1in\centerline{\titlefont #3}\abstractfont\vskip .5in\pageno=0}

\lref\ash{A.~Ashtekar, Phys.~Rev.~Lett. {\bf 57}, 2244(1986)
\semi
A.~Ashtekar, Phys.~Rev. {\bf D36}, 1587(1986)
\semi
A.~Ashtekar, {\it Lectures on non-perturbative canonical gravity,}(World
Scienti
fic, Singapore, 1991).}
%\lref\don{S.~Donaldson, Lectures. }
\lref\EH{T.\ Eguchi, A.\ J.\ Hanson, Ann.\ Phys.\ {\bf 120}, 82(1979).}
\lref\SA{S.\ Salamon, {\it Riemannian geometry and holonomy
groups}, (John Wiley \& Sons, New York, 1989).}

\Ttitle{VPI-IHEP-92-5}{hep-th/9207056}{Einstein Manifolds in Ashtekar %
Variables:}
\vskip -0.20in
\centerline{{\titlefont Explicit Examples}}
\vskip 0.5in
\centerline{Lay Nam Chang %%%
\& Chopin Soo}
\centerline{Institute for High Energy Physics}
\centerline{Virginia Polytechnic Institute and State University}
\centerline{Blacksburg, Virginia 24061-0435}
\bigskip\bigskip

\centerline{{\bf Abstract}}
\bigskip
We show that all solutions to the vacuum Einstein
field equations may be mapped to instanton configurations of the Ashtekar
variables.   These solutions are characterized by properties of the
moduli space of the instantons.   We exhibit explicit forms of these
configurations for several well-known solutions, and indicate a systematic
way to get new ones.   Some interesting examples of these new solutions are
described.

%\line{\hfil PACS: 04.20.Cv, 04.20.Jb, 11.15.Kc \hfil}

\Date{7/92}
%\draftmode
%\vfill\eject
%\bye
\newsec{Introduction}
We present some solutions to the vacuum Einstein equations
based upon the \As\ variables\ash.   These
variables are convenient for implementing the canonical description
of the Einstein field equations.  The variables are $SO(3)$ gauge fields
for Riemannian manifolds,
and we shall show that the classical solutions of the field equations
correspond to the instanton sector of the gauge fields.
Not every instanton configuration can be used to define Einstein manifolds.
In this note, we present the conditions under which this definition will
be possible, and work out some explicit examples which demonstrate the
utility of such an approach.

Ashtekar's variables\ash\
can
be obtained from the $3+1$ decomposition of the Einstein-Hilbert action,
\eqn\eh{
                    {1\over 16\pi G}{\int} { ^{4}R} \sqrt{^4 g} d^4 x
}
through a series of canonical transformations\ref\hns{%
M.~Henneaux, J.~Nelson and C.~Schonblond, Phys.Rev. {\bf D39}, 434(1989).}.
The canonical pair of variables consists of the complex Ashtekar potentials
\eqn\sua{
              A_{ia} = i K_{ia} - {1\over 2} \epsilon_{abc} \omega_{i}^{bc}
}
and the densitized triad of weight $1$,
\eqn\triad{
             \tilde{\sigma}^{ia} = \sqrt{^3 g}\sigma ^{ia}
}
The canonical variables obey the Poisson bracket relations
\eqn\pb{\eqalign{
                  \{A_{ia}(\vec x ), \tilde{\sigma}^{jb}(\vec y )\}_{PB}
             =& {i\over 2} \delta ^{j}_{i} \delta ^{b}_{a}
\delta ^3 (\vec x - \vec y ) \cr
       \{A_{ia}(\vec x ) , A_{jb}(\vec y ) \}_{PB} =& 0 \cr
      \{\tilde{\sigma}^{ia}(\vec x ) , \tilde{\sigma}^{jb} (\vec y ) \}_{PB} %
                        =& 0 \cr
}}
for all space-time points $\vec x , \vec y$ on the constant-$x^0$ 3-dimensional
hypersurface ${\cal M}^3$.    In the above, a factor of $16\pi G$ has been
suppressed on the right hand side.
For concreteness, we suppose that ${\cal M}^3$
carries the signature $+++$.  Our convention will be such that, unless
otherwise stated, lower case Latin indices run from 1 to 3, while upper
case and Greek indices run from 0 to 3.   In the above,
$\omega$ is the torsionless spin connection compatible with the triads:
\eqn\notor{
                       d\sigma ^a + \omega ^{a}_{b} \wedge \sigma ^{b} = 0
}
and modulo the constraint which generates triad rotations,
\eqn\trirot{
          K_{ia} = \sigma ^{j}\,_{a}K_{ij}
}
where $K_{ij}$ is the extrinsic curvature.

In terms of the Ashtekar variables, the constraints generating local $SO(3)$
gauge transformations, or triad rotations, which leave the spatial metric
\eqn\metric{
           g^{ij} = \sigma ^{ia}\sigma^{j}_{a}
}
invariant can be written in the form Gauss' law:
\eqn\gauss{
                   G_a \equiv {2\over i} \left( {\cal D}_i \tilde{\sigma}^i  %
                         \right)_a  \simeq 0
}
Ashtekar showed \ash\ that, modulo Gauss law constraints, the usual
``supermomentum'' and ``superhamiltonian'' constraints of ADM\ref\adm{%
R. Arnowitt, S. Deser and C. W. Misner, Phys. Rev. {\bf 116}, 1322 (1959);
Phys.
 Rev. {\bf 117}, 1595 (1960); J. Math. Phys. {\bf 1}, 434 (1960).} achieve
remarkable simplifications when expressed
in terms of the new variables.   Indeed, the ``supermomentum'' constraint
\eqn\sp{
            -2\pi_i ^{j}|_ j  \simeq 0
}
is proportional to
\eqn\hami{
        {\cal H}_i \equiv {2\over i} \tilde{\sigma}^{ia} F_{ija} \simeq 0
}
while the ``superhamiltonian'' constraint
\eqn\sham{
          {\sqrt{g}\over 16\pi G}\left( \tr  K^2 -
                     ^3R - \left(\tr  K \right)^2 \right) \simeq 0
}
is equivalent to
\eqn\ham{
            \epsilon_{abc}\tilde{\sigma}^{ia}\tilde{\sigma}^{jb}  %
                     F_{ij}^{c} \simeq 0
}
The quantity $K$ is the extrinsic curvature given by
\eqn\extcurvature{
               K^{ij} = -{16\pi G\over \sqrt{g}} \left( %
                       \pi^{ij} - {1\over 2} \pi g^{ij} \right)
}

The presence of a cosmological term
\eqn\cosmot{
    S_C = {2\lambda \over 16\pi G} \int \sqrt{^4g} d^4 x
}
in the action modifies the usual ``superhamiltonian'' constraint in
that one will need to add a new term:
$$
              {2\lambda\over 16\pi G} \sqrt{^3g}
$$
to the left hand side of \sham, and ${\cal H}$ in \ham\ becomes
\eqn\haml{
             {\cal H} = \epsilon_{abc}\tilde{\sigma}^{ia}\tilde{\sigma}^{jb}
                  F_{ij}^{c} + \Lambda \epsilon_{abc}\epsilon_{ijk}
                   \tilde{\sigma}^{ia}\tilde{\sigma}^{jb}\tilde{\sigma}^{kc}
}
with $\Lambda = \lambda /3$.

In the case of metrics with Euclidean signature, one should drop all factors
of $i$ in \sua\ and \pb, and may further assume that the Ashtekar variables
are all real.   The ``superhamiltonian'' constraint in the ADM formalism
for Euclidean signature becomes
\eqn\hame{
         {\cal H}_{\rm E} = {\sqrt{g}\over 16\pi G}
        ( -\tr K^2 - (^{3}R) + ( \tr  K )^2 + 2\lambda )  \simeq 0
}
Modulo Gauss law constraints, ${\cal H}_{\rm E}$ is still proportional
to $\epsilon_{abc}\tilde{\sigma}^{jb}F_{ija}$.  A short computation gives
\eqn\hamei{
          -16\pi G {\cal H}_{{\rm E} j} = 2 \sqrt{g} \left( K_j ^i |i
                     - K_i ^i |_j \right) = 2 \tilde{\sigma}^{ia} F_{ija}
}

\newsec{Classification Scheme for Solution Space of Constraints}
In this section, we exhibit a classification scheme of the solutions of
Ashtekar's constraints, and discuss its connection to results appearing
in the literature.

It is known that all solutions of the Einstein field equations in 4D can
be classified according to the canonical forms of the Riemann-Christofel
curvature tensor.  Such a scheme was first given by Petrov\ref\petrov{%
A.~Z.~Petrov, Doklady Akad.\ Nauk.\ SSSR {\bf 105}, 905(1955) \semi
A.~Z.~Petrov, {\it Einstein Spaces}, Pergamon Press, Oxford (1969).} and
then further extended by Penrose\ref\penrose{R.~Penrose, Ann.\ Phys.\ {\bf
10}, 171(1960).} in the context of spinors and null tetrads.

In the ADM formalism, the ``supermomentum'' and ``superhamiltonian''
constraints
are projections of the Einstein field equations tangentially and normally
to the three-dimensional hypersurface ${\cal M}^3 $, on which the initial
data compatible with the constraints is specified.  The solutions to the
constraints when stacked up according to their $x^0$-evolution by the
Hamiltonian are then the solutions to the field equations.
The natural question to ask is if a similar classification scheme can be
set up in the phase space defined by the Ashtekar variables.   What we will
do in this section is present one such scheme.

In the ADM formalism, the metric is assumed to be non-degenerate.  Ashtekar's
formulation allows for both degenerate and non-degenerate metrics.  This
is because the relevant constraints \gauss, \hami, and \haml\ do not
involve the inverse of the momenta $\tilde{\sigma}^{ia}$.  For non-degenerate
metrics, the magnetic field of the Ashtekar connection \sua,
$B^{ia} = {1\over 2}\epsilon^{ijk}F^{a}_{jk}$, can be expanded in terms of
the densitized triad $\tilde{\sigma}^{ia}$, and the most general solution
to the three-dimensional diffeomorphism constraint \hami\ is
\eqn\magi{
             B^{ia} = \tilde{\sigma}^{ib} S^{a}_{b}
}
with $S$ being a symmetric $3 \times 3$  $\vec x$-dependent matrix.  Observe
that \magi\ is a solution of the diffeomorphism constraint even for the
case of degenerate metrics.

The ``superhamiltonian'' constraint \haml\ becomes an algebraic relation:
\eqn\hamla{
            \left(\det \ts \right) \left(%
                   S^a _a + \lambda \right) = 0
}
which has the solution
\eqn\trace{
                \tr S = S^a _a = -\lambda
}
for non-degenerate metrics.   \hamla\ will not fix $\tr S$ when the metric
is degenerate.   It is intriguing to note the apparent shift in the
specification
of the dynamical degree of freedom from $\det \tilde{\sigma}$ to $\tr S$
in the case of degenerate metrics.  Metrics which become degenerate at
certain points in space-time may well be important in topology changing
situations in classical and quantum gravity.

For space-times with Lorentzian signature, the Ashtekar gauge potential
$A$ is complex,
and hence so is $S$.   Complex symmetric matrices can be
classified according to the number of independent eigenvectors and
eigenvalues, according Table 1.

Since $S_{ab}$ is gauge invariant, one may classify the matrix in terms
of the roots of its characteristic polynomial.   These can in turn be
expressed by:
\eqn\char{\eqalign{
             C_1 =& - \tr S = \lambda \cr
             C_2 =& {1\over 2}\left(C_1 \tr S + \tr {S^2} \right) \cr
             C_3 =& -{1\over 3}\left(C_2 \tr S + C_1 \tr {S^2} + \tr {S^3} %
                         \right) \cr
}}

The Bianchi identity for the magnetic field associated with $A$ further
implies the consistency condition:
\eqn\bianchi{
              \left[{\cal D}_i (S\cdot \ts ^i ) \right]_a = 0
}
or
\eqn\bianchig{
                   \ts ^{ia}\left({\cal D}_i S \right)_{ab} = 0
}
when one takes into account \gauss.

There have been attempts to obtain metric-independent gravity theories
by expressing $\ts ^i$ in terms of $B^i$ \ref\jacob{R.~Caporilla, J.~Dell,
and T.~Jacobson, Phys.\ Rev.\ Lett. {\bf 63}, 2325(1989).}.  However, in view
of the displayed classification scheme, this is not the most natural way to
proceed.  For instance, the scheme of \jacob\ will not work for
the simple $F=0$ sector, which has $S=0$ for finite
momenta.  We shall elaborate on the significance of the cases when $S$ is
degenerate later on.   When $S$ is invertible, we do obtain the results of
\jacob, with
$$
                \ts ^{ia} = \left(S^a _b \right)^{-1}B^{ib}
$$
and the constraints
\eqn\jac{\eqalign{
      ~& B^{ia}\left({\cal D}_i S \right)_{ab} =  0 \cr
      ~& S^{-1}_{ab} = S^{-1}_{ba}  \cr
     ~& \left(\det B \right) \left[ (\tr {S^{-1}})^2  - \tr \{(S^{-1})^2 \} %
            +    2\lambda \det {S^{-1}}\right] = 0 \cr
}}
As noted by the authors of \jacob, these are seven equations on the nine
complex components of $S^{-1}$, and the solutions should give the two
unconstrained field degrees of freedom associated with general relativity in
4D.  When $S$ is degenerate, though, as we will show, there could arise
phases with fewer degrees of freedom.

It should be emphasized that, as in the Petrov classification scheme,
types II, III and N do not occur for space-times with Euclidean signatures.
This is because the corresponding Ashtekar variables are all real, so that
$S$ is real and symmetric, and there are always three distinct eigenvectors.

For the case when there is only one eigenvalue, the three roots \char\
are not independent:
\eqn\one{
              \tr {S^3} = (\tr S)(\tr {S^2}) = {1\over 9}(\tr S)^3 = %
                     -{\lambda^3 \over 9}
}
When two of the eigenvalues are the same, the relationship among the roots
is
\eqn\two{
           6\left[\tr {S^3} + \lambda \tr {S^2} - {2\lambda ^3 \over 9}
\right]^
2 %
         =   \left[ \tr {S^2} - {\lambda ^2 \over 3} \right]^3
}

The initial value data thus falls into distinct classes with strikingly
distinct properties.  For instance, type I has three $\vec x$-dependent
eigenvalues for $S$, whose sum is restricted to $-\lambda$, while for type O
one has only one $\vec x$-independent eigenvalue $-\lambda /3$.  This
mismatch in the allowable fluctuations is highly suggestive of distinct
phases in the theory.  For example,
we may show\ref\csbrst{L.\ N.\ Chang and C.\ Soo, {\it Ashtekar variables and
the topological phase of quantum gravity,} in Proceedings of the XXth. DGM
Conference, eds. S. Catto and A. Rocha, (World Scientific, Singapore, 1991)
\semi {\it BRST cohomolgy and
invariants of 4D gravity in Ashtekar variables,} VPI-IHEP -92-4,
hepth@xxx/9203014.}
that type O ($S_{ab} = -(\lambda /3) %
\delta_{ab}$) can be identified with an unbroken topological quantum field
theory (TQFT), describing a topological phase in quantum gravity.

The classification scheme described so far becomes equivalent to the usual
Petrov classification for non-degenerate metrics.   In this case,
\eqn\nondeg{
            S_{ab} = R_{\widetilde{0a} 0b}  - R_{0a0b}
}
%%%%%%%

\newsec{Equations of Motion and Anti-Instantons}
In this section, we exhibit the manifestly covariant equations of
motion for the \As\ variables and discuss the implications.  We choose
to work explicitly with metrics of Euclidean signature and use $SO(3)$
instead of $SU(2)$ gauge potentials,  but we will indicate the necessary
modifications
for metrics of Lorentzian signature.    It will become clear as we go
along, that there are Einstein manifolds that cannot be described globally
by $SU(2)$ \As\ potentials, but can be described by $SO(3)$ connections.
This has to do with the fact that not all $SO(3)$ connections can be lifted
to be $SU(2)$ connections with integer second Chern class, but {\it all}
$SU(2)$ connections can be thought of as $SO(3)$ connections with the
first Pontrjagin class being a multiple of four.  We will furnish examples
of such manifolds below.

In working with metrics of Euclidean signature, we should drop all $i$'s,
starting with Eqn.~(1.1).  In the spatial gauge, we have
\eqn\kk{
          e_{A\mu} = \left(\matrix{N & 0 \cr
                                  e_{aj}N^j  & e_{ai} \cr} \right)
}
Eqn.~\kk\ in no way compromises the values of the lapse and shift functions
$(N, N^j )$\adm, and is compatible with the ADM decomposition of the metric:
\eqn\shift{\eqalign{
     ds^2 =& \mp e_{0}\,^{2} + e_{1}\,^{2} + e_{2}\,^{2} + e_{3}\,^{2} \cr
     ~~~  =& \mp N^2 (dx^{0})^2 + g_{ij}\left(dx^1 + N^idx^0 \right)
             \left(dx^j + N^j dx^0 \right)  \cr
}}
where $e_A$ is the $1$-form $e_{A\mu}dx^{\mu}$ and the $+(-)$ sign is to
be used for metrics of Euclidean (Lorentzian) signature.  On the constant
$x^0$-hypersurface ${\cal M}^3$, $e_0$ vanishes, and we may write
\eqn\mag{
            F_a = e^0 \wedge T_{ab}e^b + {1\over 2} S_{a}^{b} %
                \epsilon_{bcd} e^c \wedge e^d
}
which gives Eqn.~(2.1).
$T_{ab}$ however must be chosen carefully because Eqn.~\mag\ implies
that on ${\cal M}^3$
\eqn\magcomp{\eqalign{
           F_{0ia} =&  T_{ab}\left(e^0\,_0 e^b\,_i  - e^{0}\,_{i}e^{b}\,_{0}
\right) + S_{ab}\epsilon^{bcd}e_{c0}e_{di}
                             \cr
         ~~~       =&  NT_{ab}e^{b}\,_{i} + S_{ab}\epsilon^{bcd}e_{c0}e_{di}
}}
For Riemannian manifolds, apart from a boundary term that does not
contribute to the equations of motion, the Hamiltonian in the \As\ formalism
is$[1]$:
\eqn\hamr{\eqalign{
   H =& \int_{{\cal M}^3} d^3 x \ut{N} \left(\epsilon_{abc}\ts ^{ia}%
            \ts ^{jb} F_{ij}^{c} + ({\lambda \over 3})\epsilon_{abc}
              \epsilon_{ijk} \ts^{ia}\ts^{jb}\ts^{kc}\right)   \cr
   ~~~ & + 2 N^i \ts^{ja}F_{ija} - 2 A_{0a}({\cal D}_i \ts^{i} )^a  \cr
}}
and the evolution equation for $A_{ia}$ on $\cM^3$
gives
\eqn\evolv{\eqalign{
       \dot{A}_{ia} =&  \left\{A_{ia}, H \right\}_{\rm PB} \cr
       ~~~          =&  \ut{N} \eps_{abc}\ts^{jb}F_{ij}^{c} + {1\over 2} %%
                      \lambda \ut{N} \eps_{ijk}\eps_{abc}\ts^{jb}\ts^{kc} \cr
         ~~~         & + \partial_i A_{0a}  - \eps_{a}~^{bc}A_{0b}A_{ic} - %%
                       N^j F_{ija}  \cr
}}
With the use of Eqn.~\magi\ and assuming non-degenerate metrics, we can rewrite
Eqn.~\magcomp\ as
\eqn\nmagcomp{\eqalign{
          F_{0ia} =& -N S_{ab}e^{b}~_{i} + N e_{ia}\left(S^{c}\,_{c}  + \lambda
                          \right)  \cr
             ~~~   & + S_{ab}\eps^{bcd}e_{c0}e_{di} \cr
}}
The second term vanishes because of the ``superhamiltonian'' constraint and
comparing with Eqn.~\magcomp\ we observe that the consistent choice
for $T_{ab}$ is:
\eqn\choice{
           T_{ab} = -S_{ab}
}
Thus we have
\eqn\fmagcomp{
           F_{a} = S_{ab}\left(-e^0 \wedge e^b + {1\over 2}\eps^{b}\,_{cd}
                     e^c \wedge e^d \right)
}

Similarly, for the evolution of $\ts^{ia}$, we have
\eqn\evolvsig{\eqalign{
       \dot{\ts}^{ia} =& \left\{\ts^{ia}, H \right\}_{\rm PB} \cr
             ~~~      =&  \eps^{a}\,_{bc} \left[{\cD}_j (\ut{N}\ts^i)\right]^b
                          \ts^{jc} + \eps^{a}\,_{bc}\ut{N}\ts^{ic}({\cD}_j
\ts^{
j})^b  \cr
             ~~~       & + \left[{\cD}_j (N^j \ts^{i})\right]^a - (\partial_j
N^
i )
                           \ts^{ja}  \cr
             ~~~       & -N ({\cD}_j \ts^j )^a  + A_{0c}\eps^{a}\,_{b}\,^{c} %%
                        \ts^{ib}    \cr
}}
It is not difficult to show that the equations of motion for the \As\ variables
can then be succinctly written as
\eqna\aseqn
$$\eqalignno{
%\eqn\aseqn{\eqalign{
            F_a =&  S_{ab} \Sigma^b &\aseqn a\cr
           \left(D\Sigma \right)^a =&  0  &\aseqn b\cr
}$$
with
\eqna\ascond
%\eqn\ascond{\eqalign{
$$\eqalignno{
           S_{ab} =& S_{ba}  &\ascond a\cr
          \Tr S  =& - \lambda &\ascond b\cr
}$$
Here,
\eqn\SSigma{
           \Sigma^a \equiv -e^0 \wedge e^a + {1\over 2}\eps^{a}\,_{bc} %
                      e^b \wedge e^c
}
and $D$ is the covariant derivative with respect to the \As\ connection 1-form.
The nine $x^0$-evolution equaitons for $A_{ia}$ are contained in
Eqn.~\aseqn{a}\
while the twelve equations in Eqn.~\aseqn{b} can be split off into the set of
three equations:
\eqn\starsig{
                  ^{\ast}\left[\left(D\Sigma\right)^a |_{\cM ^3 } \right]  %
                                  = 0
}
which is equivalent to the set of
Gauss Law constraints, and the nine equations:
\eqn\starsigg{
                 \left[^{\ast}\left(D\Sigma\right)^a \right] |_{\cM ^3 } = 0
}
which, modulo the Gauss Law constraints, are equivalent to the $x^0$-
evolution equations for $\ts ^{ia}$, Eqn.~\evolvsig.  Ashtekar's transcription
of the ``supermomentum'' and ``superhamiltonian'' constraints of general
relativity takes the simple form of \ascond{a,b}.
(See also \ref\JS{R. Capovilla, J. Dell, T. Jacobson and L. Mason, Class. %
Quantun. Grav. {\bf 8}, 41 (1991)\semi R. Capovilla, T. Jacobson and J. Dell,
i
bid. {\bf 7} L1, (1990).}
for an alternative derivation
of the equations of motion using self-dual two-forms as fundamental variables
and a discussion of gravitational instantons as $SU(2)$ rather than
$SO(3)$ gauge fields.)

We shall now examine the meaning of the equations of motion.
Firstly, observe that $\Sigma ^a$ is explicitly anti-self-dual:
\eqn\asds{
                 ^{\ast}\Sigma ^a = - \Sigma ^a
}
Since $F_a$ is the product of a zero form $S$ with the 2-form $\Sigma$,
Eqn.~\asds\ implies that
\eqn\asdf{
             ^{\ast}F_a = - F_a
}
As a result, all Einstein manifolds correspond to anti-instantons of the \As\
potentials.  However, the converse is not always true.  In general, the
curvature of an arbitrary anti-instanton can be expanded in terms of $\Sigma^a$
via
\eqn\ainst{
        F_a = Y_{ab}\Sigma ^b
}
But the quantity $Y$ will have to satisfy Eqn.~\ascond\ and Eqn.~\aseqn{b}\
before the anti-instanton can correspond to an Einstein manifold.

The twelve equations in Eqn.~\aseqn{b}\ suggest that the 1-form $A_a$
can be expressed in terms of the vierbein $e_A$.  This is indeed
the case, for the solution to Eqn.~\aseqn{b}\ is precisely
\eqn\aspot{
       A_a = \omega_{0a} - {1\over 2} \eps_{a}\,^{bc} \omega_{bc}
}
where $\omega_{AB}$
can be determined uniquely from $e_A$ through
\eqn\torsion{
            de_A + \omega_{AB} \wedge e^B  = 0
}
Eqn.~\aspot\ says that, apart from a factor of 2, $A_a$ is the anti-self-dual
part of the spin-connection and so the curvature 2-form of $A_a$ can
be expressed as
\eqn\fspin{
        F_a = R_{0a} - {1\over 2}\eps_{a}\,^{bc}R_{bc}
}
where $R_{AB}$ is the curvature 2-form of the spin-connection.  It is then
not difficult to show that Eqn.~\aseqn{b}\ is satisfied
if and only if
\eqn\sdual{
      S_{ab} = R_{\widetilde{0a}0b} - R_{0a0b} = R_{0a\widetilde{0b}} %
                       - R_{\widetilde{0a}0b}
}
and so the constraints Eqn.~\ascond\ imply that
\eqn\rdual{
           R_{ABCD} = R_{\widetilde{AB}\widetilde{CD}}
}
and the Ricci scalar becomes
\eqn\ricci{
             R = 4\lambda
}
These equations are completely equivalent to the pure gravity field equations
defining Einstein manifolds.

Dimension four is the lowest dimension for which the Riemann curvature tensor
assumes its full complexity.  It is also the dimension which has the
peculiarity that the curvature 2-form can be decomposed into parts taking
values in the $(\pm)$ eigenspaces $\Lambda_2 ^{\pm}$ of the Hodge duality
operator.  The Riemann curvature tensor, having four indices, can be dualized
on the left or on the right, so that it can be viewed as a $6 \times 6$ mapping
of $\Lambda^2_{\pm} \to \Lambda^2 _{\pm}$\ref\atiyah{M. F. Atiyah, N. Hitchin
an
d I. M. Singer, Proc. Roy. Soc. Lond. {\bf A362}, 425 (1978).}:
\eqn\map{
         \left(\matrix{{\cal A} & {\cal C}^{+} \cr
                       {\cal C}^{-} & {\cal B} \cr} \right)
}
where in components,
\eqn\mapcomp{
          {\cal A}_{ab} \equiv +\left(R_{0a0b} + R_{0a\widetilde{0b}}\right)
                    + \left(R_{\widetilde{0a}0b} +
R_{\widetilde{0a}\widetilde{0
b}}%
                   \right)
}
and ${\cal B}$ and ${\cal C}$ are defined similarly according to the
signs of the following:
\eqn\mapcomps{
          {\cal A} \sim  \left(+,+,+,+\right),
          {\cal B} \sim  \left(+,-,-,-\right),
          {\cal C}^{+} \sim  \left(+,-,+,-\right),
          {\cal C}^{-} \sim  \left(+,+,-,+\right)
}
It is easy to check that ${\cal A} ({\cal B})$ is self-dual (anti-self-dual)
with respect to both left and right duality operations, while ${\cal C}^{+}%
({\cal C}^{-})$ is self-dual (anti-self-dual) under left duality
and anti-self-dual (self-dual) under
right duality operations.  A metric is Einstein if and only if
${\cal C}^{\pm} = 0$, {\it i.e.} when Eqn.~\map\ assumes a block
diagonal form.  In view of Eqns.~\asdf\ and \fspin, $F_a$ is the doubly
anti-self-dual part of the curvature and apart from a multiplicative factor,
$S$, can be identified with ${\cal B}$ when the equations of motion are
satisfied.  In this context, for Einstein manifolds, the \As\ formulation
is the realization of Proposition 2.2 of \atiyah\ in the
canonical framework.  However, it should be emphasized that it is the
remarkable
simplification of the constraints provided by \As\ that makes the
non-perturbative quantization scheme viable.   While it appears that only
half of the non-vanishing components of the Riemann curvature tensor is
contained
in $F_a$, the equations of motion are completely equivalent to Einstein's
field equations for non-degenerate metrics.   Actually, ${\cal A}$ and
${\cal B}$ interchange under a reversal of orientation because a reversal
of orientation changes the definition of self- and anti-self-duality.

While not all Einstein manifolds have anti-self-dual Riemann or Weyl
tensors, a manifold is Einstein only if the curvature tensor constructed from
the anti-self-dual part of the spin connection is anti-self-dual.  It is
precisely this property which allows for the description of all Einstein
manifolds in terms of anti-instantons of the \As\ variables.

As a corollary, we note that for Einstein manifolds, the Weyl 2-form is
\eqn\weyl{
    W_{AB} = R_{AB} - {\lambda\over 3} e_A \wedge e_B
}
so the anti-self-dual part of the Weyl 2-form $W^{-}_a$
becomes
\eqn\asdw{\eqalign{
    W^{-}_a =& R_{0a} - {1\over 2} \eps_{a}\,^{bc}R_{bc} +
{\lambda\over 3}\left[-e_0 \wedge e_a + {1\over 2}\eps_{a}~^{bc}e_b
\wedge e_c \right]
\cr
           =& F_a + {\lambda\over 3}\Sigma_a \cr
}}
so an Einstein manifold is conformally flat or self-dual (half-flat when
$\lambda = 0$) if and only if
\eqn\cflat{
            F_a = -{\lambda\over 3}\Sigma_a
}
or
\eqn\cflats{
             S_{ab} = -{\lambda\over 3} \delta_{ab}
}
According to our classification, this situation corresponds
precisely to type $O$.

It is possible to eliminate $S_{ab}$ from the equations of motion.  We have
\eqn\eqnos{
          \Sigma^a \wedge \Sigma^b = -2\delta^{ab}\left(\ast 1\right)
}
where $(\ast 1)$ is the 4-volume element.  So from the equations of
motion Eqn.~\aseqn\
\eqn\sf{
           S_{ab} = -{1\over 4}\ast \left(F_a \w\Sigma_b + %%
                     \Sigma_a \w F_b \right)
}
and the equations of motion can be written as
\eqna\eqmot
$$\eqalignno{
          F_a =& -{1\over 2}\left[^{\ast}\left(F_a \wedge \Sigma_b \right) %
                    \right] \Sigma^b &\eqmot a\cr
         \left(D\Sigma\right)^a =& 0 &\eqmot b\cr
         \eps_{a}\,^{bc}F_b \wedge \Sigma_c =& 0 &\eqmot c\cr
          F_a \wedge \Sigma_a =& -2\lambda \left(\ast 1 \right) &\eqmot d\cr
}$$

%\bye

\newsec{Invariants and the Ashtekar variables}
Unlike other fields, the gravitational field describes the dynamics of
space-time.  Any viable classical and quantum theory of the gravitational field
must therefore be able to take into account not just the local description of
curvature, but also the large scale global and topological aspects of the
structure of space-time.  We shall see how the \As\ variables can be used to
capture the global invariants in 4D,
especially those associated with Einstein manifolds.

As we have discussed in section 2, a specification of the initial value data
is equivalent to a specification of the characteristic classes of $S$ which
is compatible with the constraints. We may take the gauge-invariant quantities
on $\cM ^3$ to be $\Tr S = -\lambda$, $\Tr S^2$, and $\Tr S^3$, from which
we can reconstruct the characteristic classes of $S$.   Their integrals over
$\cM ^3$ should reflect global properties of $\cM ^3$.

It is not difficult to show that when the equations of motion are satisfied,
\eqna\invs
$$\eqalignno{
      \Tr S =& -\lambda  &\invs a\cr
     \Tr S^2  =& {1\over 8}\left\{\left(R_{\widetilde{AB}\widetilde{CD}} -
                    R_{AB\widetilde{CD}}\right)R^{ABCD}\right\} &\invs b\cr
    \Tr S^3  =& -{1\over 16} \left\{\left(R_{ABCD} -
R_{AB\widetilde{CD}}\right)
                             R^{CDEF}R_{EF}\,^{AB} \right\} &\invs c\cr
}$$
Thus their integrals over compact, closed 4-manifolds $\cMM$ give
\eqn\intinva{
     \int_{\cMM } \left(\Tr S \right) = -\lambda V = -{\lambda\over 6}
                       \int \Sigma ^a \wedge \Sigma_a
}
where $V$ is the volume of $\cMM$, and
\eqn\intinvb{\eqalign{
       \int_{\cMM} \left(\Tr S^2 \right) =& 2\pi^2 \left(2\chi (\cMM )
                          - 3\tau (\cMM )\right)  \cr
                            =& -{1\over 2} \int F_a \wedge F^a   \cr
                           =& -2\pi ^2 P_1 \cr
}}
where $\chi (\cMM)$ and $\tau (\cMM )$ are the Euler characteristic
and signature of $\cMM$, while $P_1$ is the Pontrjagin number of the
$SO(3)$ \As\ connection.
Finally,
\eqn\intinvc{
        \int_{\cMM} (\Tr S^3 ) = -{1\over 2} \int_{\cMM} S_{ab}F^a \wedge F^b
}

Observe that the signature $\tau (\cMM)$ depends on the orientation of
$\cMM$.  Indeed,
\eqn\rev{\eqalign{
         \tau (\cMM) =& {\rm dim}\; H^2 _{+} \; - \; {\rm dim}\; H^2 _{-} \cr
                     =& b_2^{+} - b_2^{-}  \cr
}}
where $H^2 _{\pm}$ are the self-dual and anti-self-dual subspaces of the
second cohomology group, and $b_2^{\pm}$ are the corresponding Betti numbers.
Reversing the orientation interchanges self-dual and anti-self-dual 2-forms,
so that
\eqn\revsig{
         \tau (\overline{\cMM}) = - \tau (\cMM)
}
where $\overline{\cMM}$ has the opposite orientation relative to $\cMM$.
Reversing the orientation changes the spin connections in general, and
thus the \As\ connections via Eqn.~\aspot.    For example, consider
$$
          de^A = -\omega ^A_B \wedge e^B
$$
A transformation of the form $(e^0 , e^a ) \to (-e^0 , e^a )$ reverses
the orientation, though it does not change the metric $ds^2$.  The new
spin connections become
\eqn\spinr{
            \omega_{0a} \to -\omega_{0a}; \qquad \omega_{ab} \to \omega_{ab}
}
so that the \As\ connections transform as
\eqn\asr{\eqalign{
            A_a =& \omega_{0a} - {1\over 2}\eps_{a}~^{bc}\omega_{bc} \cr
                 & \to A_a - 2\omega_{0a} \cr
}}
The Pontrjagin numbers of the \As\ connections with respect to the two
different
orientations are
\eqn\pontrj{\eqalign{
         P_1 ^{+} =& 3\tau(\cMM) - 2\chi(\cMM)  \cr
         P_1 ^{-} =& 3\tau(\overline{\cMM}) - 2\chi(\overline{\cMM}) = -3\tau(
\cMM) - 2\chi(\cMM)  \cr }
}
Since $P_1 ^{\pm}$ are the Pontrjagin numbers of the anti-self-dual \As\
connections,
$$
            P_1^{\pm} \leq 0
$$
An immediate consequence is the Hitchin bound for compact, closed Einstein
manifolds\ref\hitchin{H.\ Hitchin, J.\ Diff.\ Geom.\ {\bf 9}, 435, (1974).}
\eqn\hitc{
               |\tau | \leq {2\over 3} \chi
}
For compact, closed Einstein manifolds with Euclidean signatures,
$$\eqalign{
         \chi({\cMM}) =& {1\over 32\pi^2 } \int
R_{\widetilde{AB}\widetilde{CD}}
                        R^{ABCD} \cr
                    =& {1\over 32\pi^2 } \int \left(R_{ABCD}\right)^2  \geq 0 %
                        \cr
}$$ with the equality holding only if $\cMM$ is flat. Moreover $\tau(\cMM)$ and
$\chi(\cMM)$ can be computed from the $SO(3)$ \As\
connections through
\eqna\sign
$$\eqalignno{
           \tau(\cMM) =& {1\over 6} \left(P_1^{+} - P_1^{-} \right) &\sign a\cr
          \chi(\cMM) =& -{1\over 4} \left(P_1 ^{+} + P_1^{-}\right) &\sign b\cr
}$$

If the Einstein manifold possesses an orientation reversing diffeomorphism,
then $P_1 ^{+} = P_1^{-}$, and $\tau = 0$.   The vanishing or non-vanishing
of the signature has important physical implications.  For according to
the index theorem for the spin complex for closed, compact Riemannian
manifolds,
\eqn\spin{\eqalign{
             n_+ - n_- =& -{1\over 24} P_1\left(T(\cMM)\right) \cr
                       =& -{1\over 8} \tau (\cMM)  \cr
}}
where $n_{\pm}$ are the number of $\pm 1$ chirality zero-frequency solutions
of the Dirac equation.   $P_1\left(T(\cMM)\right)$ is the Pontrjagin number
of the tangent bundle, {\it i.e.} of the $SO(4)$ spin connection, and
is related to the $\tau (\cMM)$ by the Hirzebruch signature theorem:
$$
        P_1 \left(T(\cMM)\right) = 3 \tau (\cMM )
$$
Thus
$$
         \tau (\cMM ) = 0 \qquad {\rm mod} ~~8
$$
for spin manifolds, since $n_+ - n_-$ must be an integer.  An orientable
manifold ($\cW _1 =0$) has a spin structure iff $\cW _2 = 0$.   Here
$\cW $ refer to the Stiefel-Whitney class.  A simply-connected,
compact, closed manifold of dimension four has a spin structure
iff its intersection form is even, and this spin structure is
unique\ref\donkron
{%
S.\ K.\ Donaldson and P.\ B.\ Kronheimer, {\it The geometry of four-manifolds},
(Oxford Mathematical Monographs, Clarendon Press, Oxford, 1990).}.
Actually, for the case of simply-connected, compact, closed, {\it smooth}
four-manifolds, the intersection form, and hence the topology via Freedman's
theorem, is determined by $\tau$ and $\chi$, and whether the intersection
form is even (i.e. $\cW _2 = 0$) or odd. This can be explained as
follows: Indefinite
intersection forms are determined by their rank, signature, and type (even
or odd).
The rank of the intersection form is the second Betti number.   But
\eqn\btwo{
      b_2 = b_2^{+} + b_2^{-} = \chi - 2
}
for simply-connected, compact, closed four-manifolds. $\tau$ is the signature
of
 the intersection form.
Although
there are many definite intersection forms of the same rank and signature,
Donaldson's theorem \ref\donaldth{S.\ K.\ Donaldson, J.\ Diff.\ Geom.\ {\bf
18},
269, (1983).} asserts that differentiable four-manifolds
with definite intersection forms must be of the standard type
%$\pm \bigoplus^{n}_{ }(1)$.
${\rlap{\raise2.5ex\hbox{$~\, n$}}\bigoplus{}}\llap{%
\lower2.5ex\hbox{$\pm ~$}} (1)$.
So specification of
$P_1 ^{\pm}$ and whether the manifold is spin ($\cW _2 =0$) or not corresponds
to a complete specification of the intersection form of a smooth,
simply-connected, compact, closed four-manifold.  Freedman's theorem
\ref\freedth{M.\ Freedman, J.\ Diff.\ Geom.\ {\bf 17}, 1357, (1982).} asserts
that given an even (odd) intersection form, there is exactly one (two,
distinguished by their ${\cal Z}_2$-valued Kirby-Siebenmann invariant)
simply-connected, closed, compact, topological four-manifold representing
that form.

Before we proceed to specific illustrations, we remark that the
third invariant Eqn.~\invs{c}, which involves the explicit form of $S$ could
provide a new differential invariant for Einstein manifolds, since the
intersection form has already been accounted for by Eqn.~\intinvb,
at least for the case when they are
smooth, simply-connected, closed, and compact. See also \csbrst\ for
a discussion of BRST-invariants of four-dimensional
gravity in Ashtekar variables.
\vfil\eject
%\bye

\newsec{Examples of Einstein manifolds in Ashtekar variables}
\medskip
\centerline{%
{\bf A. Known Solutions} }
\bigskip
The formalism developed in the previous sections provides a coherent
framework to discuss explicit Einstein manifolds in the context of \As\
variables.    Every known solution of the Einstein field equations
\eqn\einf{
            R_{\mu\nu}  = \lambda g_{\mu\nu}
}
can be put in the form of Eqns.~(3.11).   In fact, when the field equations
are satisfied, we can use Eqns.~(3.12) to obtain the \As\ connection, and
compute $S$ via Eqn.~\aseqn{a}.

It will be convenient to introduce the 1-forms $\Theta_a $, where $\Phi_a =
-2\,\Theta_a$ obeys the Maurer-Cartan equation for $SO(3)$:
\eqn\maurer{
            d\Phi_a + \ha \eps_{a}\,^{bc}\Phi_b \wedge \Phi_c = 0
}
We can choose the four-dimensional polar coordinates as $(R, \theta, \phi,
\psi)
$,
where for fixed $R$, $0\leq \theta \leq \pi$, $0\leq \phi < 2\pi$,
and $0\leq \psi < 4\pi$.  Next introduce
\eqn\twist{\eqalign{
           x^1 + i x^2 =& R \cos ({\theta /2}) \exp {{i\over 2}(\psi + \phi)}
\cr
           x^3 + i x^0 =& R \sin ({\theta /2}) \exp {{i\over 2}(\psi - \phi)}
\cr
}}
Then $\Theta_a$ can be written in terms of the Euler angles $\theta, \phi,
\psi$ on $S^3 $ as
\eqn\sthree{\eqalign{
           \Theta _1 =& \ha \left(\sin \psi d\theta  %%
                        - \sin \theta \cos \psi d\phi\right) \cr
          \Theta _2 =& \ha \left(-\cos\psi d\theta  %%
                       - \sin \theta \sin \psi d\phi \right) \cr
         \Theta_3 =& \ha\left(d\psi + \cos\theta d\phi\right) \cr
}}

We concentrate first on solutions with $S_{ab} = -(\lambda /3)\delta_{ab}$.
As we have explained, these solutions correspond to the conformally self-dual
sector of Einstein manifolds.   It is known that for $\lambda >0$, $S^4$ and
$CP_2$ are the only compact, closed, simply-connected four-manifolds which
are conformally self-dual\SA.
\vfil\eject

\line{%
(a)$\underline{{\bf S^{4}~ {\rm with~ the~ de~ Sitter~ metric}}}$ \hfil}
The metric for this space is given by
\eqn\deS{
    ds^2 = \left[1 + \left({R\over a}\right)^{2}\right]^{-2} %%
                \left[dR^2 + R^2 \left(\Theta_1 ~^2  + \Theta_2 ~^2 +  %%
                  \Theta_3 ~^2 \right)\right]
}
while the vierbein is expressed as
\eqn\deSv{
     e_A  = \left\{{dR\over \left(1 + \left({R\over a}\right)^2 \right)},
               {R\Theta_a \over \left(1 + \left({R\over a}\right)^2\right)} %
              \right\}
}
The corresponding \As\ connections then have the form:
\eqn\deSas{\eqalign{
           A_a  =& \omega_{0a} - \ha \eps_{a}\,^{bc}\omega_{bc} \cr
                =& -{2\Theta_a \over \left(1+ \left({R\over a}\right)^2
\right)%
                    }
}}
giving
\eqn\deSf{\eqalign{
          F_a =& dA_a + \ha \eps_{a}~^{bc}A_b \w A_c \cr
              =& -{4\over a^2} \left( - e_0 \w e_a + \ha \eps_{a}\,^{bc} %%
                        e_b \w e_c \right) \cr
}}
Thus,
$$
             F_a = S_{ab}\Sigma^b
$$
with
\eqn\deSS{
        S_{ab} = -{4\over a^2}\delta_{ab} = - {\lambda \over 3}\delta_{ab}
}
so that
\eqn\deSc{
            \lambda = {12\over a^2} > 0
}
and the diameter of the four sphere is related to $\lambda$ by
\eqn\deSr{
             a = \sqrt{{12\over \lambda}}
}

Suppose that we now reverse the orientation, by for example defining
the vierbein field to be
\eqn\deSrv{
    \overline{e}_A  = \left\{-{dR\over \left(1 + \left({R\over a}\right)^2
\right)},  {R\Theta_a \over \left(1 + \left({R\over a}\right)^2\right)}
              \right\}
}
The \As\ connections then change to the form:
\eqn\deSra{
         \overline{A}_{a}  = -{2\over a^2}{R^2 \Theta_a \over \left(1+ \left(
{R\over a}\right)^2 \right)}
}
although
\eqn\deSrf{
            \overline{F}_a = -{4\over a^2}\overline{\Sigma}_{a}
}
so that $S$ is unchanged.   $S^4$  has an orientation reversing diffeomorphism.
By using the explicit form for the \As\ connections, one obtains that
the first Pontrjagin number equals $-4$, and is preserved under the
reversal.
Thus Eqns.~\sign{a}\ and \sign{b}\ yield
\eqna\deSsign
$$\eqalignno{
            \tau\left(S^4 \right) =& 0  &\deSsign   a\cr
            \chi\left(S^4 \right) =& 2 &\deSsign  b\cr
}$$
Note that for $S^4$, the $SO(3)$ \As\ connections give $P_{1}^{\pm} = 0 \;
{\rm mod}\; 4$, and so can be lifted to an $SU(2)$ connection, with
second Chern class
\eqn\deSch{\eqalign{
          c_2 =& -{P_1\over 4}  \cr
              =& 1  \cr
}}
Actually, the \As\ connections given by Eqns.~\deSas\ and \deSra\ are
precisely the BPST (anti-)instanton solutions\ref\bpst{A. A. Belavin, A. M.
Poly
akov, A.S. Schwarz and Yu. I. Tyupkin, Phys. Lett. {\bf 59B}, 85 (1975).
\semi see also J.\ Samuel, Class. Quantum. Grav. {\bf 5}, L123 (1988).}.
Since the intersection form has rank
\eqn\intform{
          {\rm rank}\; (Q) = b_2 = \chi - 2 = 0
}
$S^4$ has $Q=\emptyset$.
% The Poincar\'e conjecture for dimension four says
%that $S^4$ is the only topologically closed compact simply-connected
%manifold having $Q=\emptyset$.

The dimension of the moduli space for a single anti-instanton on $S^4$
is known to be five\atiyah.  The parameters correspond to the size
and location of the (anti-)instanton.   For the \As\ connections, however,
diffeomorphism invariance collapses this space entirely, since the solution
must now be translationally invariant, and the size is fixed by the
cosmological constant, according to Eqn.~\deSr.  $S^4$ is not only
conformally self-dual, but it is also conformally flat.   That this is so
is also evident in the \As\ context because $S_{ab} = \lambda /3 \delta_{ab}$
implies, by Eqn.~\asdw\ that $\cW _{a}^{-} = 0$.  But $S$ is unchanged
by orientation reversal, so $\cW_{a}^{+} = 0$ also.    Hence
$\cW_{a}^{\pm} =0$, and $S^4$ is conformally flat.

\vfil\eject
%\bye
\line{%
(b)\underbar{CP$_2$ and the Fubini-Study Metric} \hfil}
\medskip
The two dimensional complex projective space is described by the Fubini-Study
metric:
\eqn\fs{
       ds^2 = {dR^2\over \left(1 + {\lambda\over 6}R^2 \right)^2} +  %
           {\left(R\Theta_1\right)^2\over\left(1 + {\lambda\over 6}R^2\right)}
+
         {\left(R\Theta_2\right)^2\over \left(1 + {\lambda\over 6}R^2\right)} +
   {\left(R\Theta_3\right)^2\over \left(1 + {\lambda\over 6}R^2\right)^2}
}
We may choose the vierbeins as
\eqn\fsv{
  e_A = \left\{ {dR\over \left( 1+{\lambda\over 6}R^2\right)}, %%
             {R\Theta_1\over \left(1 +{\lambda\over 6}R^2\right)^{\ha}} , %%
   {R\Theta_2\over \left(1 + {\lambda\over 6}R^2\right)^{\ha}}, %%
   {R\Theta_3\over \left(1 + {\lambda\over 6}R^2\right)}\right\}
}
in which case the \As\ variables are:
$$
    A_1  = {-2\Theta_1\over \left(1 + {\lambda\over 6}R^2\right)^{\ha}}
   \qquad\quad A_2 = {-2\Theta_2\over \left(1 + {\lambda\over
6}R^2\right)^{\ha}
}, %%
$$
and
\eqn\fsa{
      A_3 = {\left(-2 - {\lambda \over 6}R^2 \right) \Theta_3 \over \left(1 +
{\lambda \over 6}R^2 \right)}
}

These equations yield $F_a = S_{ab}\Sigma^{ab}$, with $S_{ab} = -(\lambda /3) %
\delta_{ab}$.  The solution is therefore again of Type O.  However, the
Pontrjagin index is found to equal
\eqn\pontcp{
      P_1 = {1\over 4\pi}\int F_a \w F_a = -3
}
As a result, the \As\ connections cannot be realized in a globally well-defined
manner as an $SU(2)$ gauge potential.

Like $S^4$, $CP_2$ is conformally flat, since $S$ is of Type O, but
unlike $S^4$, it does not have an orientation reversing diffeomorphism.
Under a reversal, we obtain $\overline{CP}_2$, which is described by the {\it
same} metric, but the vierbein becomes $(-e_0, e_a)$.   In which case,
the \As\ potentials become
$$
    \overline{A}_1 = \overline{A}_2 = 0
$$
while
\eqn\fsbara{
    \overline{A}_3 = -{\lambda R^2\,\Theta_3\over 2\left(1 + {\lambda\over
6}R^2
               \right)}
}
giving
\eqn\fsbarf{\eqalign{
       \overline{F}_1 &= \overline{F}_2 = 0 \cr
        \overline{F}_3 &= d\overline{A}_3 \cr
             &= -\lambda\left( e_0 \w e_3 + e_1 \w e_2 \right) \cr
             &= - \lambda \overline{\Sigma}_3 \cr
}}
Thus $CP_2$ is described by \As\ potentials of a non-abelian
anti-instanton, whereas $\overline{CP}_2$ is described by those of an abelian
anti-instanton.
The corresponding Pontrjagin index is found to be
\eqn\fsbarp{
           \overline{P}_1 = -9
}
which is different from that of Eqn.\pontcp.  Accordingly, the Euler
characteristic and signature are given by
\eqn\eul{
        \chi (CP_2 ) = \chi (\overline{CP}_2 ) = 3
}
while
\eqn\sig{
       \tau (CP_2 ) = - \tau (\overline{CP}_2 ) = 1
}

{}From previous studies \donkron ,
we already know
that $CP_2$ cannot support abelian instantons, while $\overline{CP}_2$ can
support only one such object.   The \As\ potential is simply that
unique abelian anti-instanton.

The matrix $S_{ab}$ for $\overline{CP}_2$ is of the form
\eqn\fsbars{
     \overline{S} = \hbox{diag}\;\left(0,\, 0,\, -\lambda\right)
}
and so the solution is of Type D.

This example shows how the \As\ variables provide a more natural
context in which to study the topological and differential invariants of
a 4-manifold.

%\bye
\bigskip
\line{%
(c)\underbar{The Schwarzschild-de Sitter solution} \hfil}
\medskip
The Schwarzschild-de Sitter metric in Euclidean space is given by
\eqn\schmet{
   ds^2 = \left(1 - {2M\over r} - {\lambda\over 3}r^2\right)d\tau^2 + %%
         \left(1 - {2M\over r} - {\lambda\over 3}r^2\right)^{-1} dr^2 + %%
    r^2d\theta^2 + r^2\sin^{2}\theta\,d\phi^2
}
where $M$ is $G/c^2$ times the mass.  Taking the vierbein fields to be
\eqn\schvier{
   e_A = \left\{\left(1 - {2M\over r} - {\lambda\over 3}r^2\right)^{\ha} %%
           d\tau, \; \left(1 - {2M\over r} - {\lambda\over 3}r^2\right)^{-\ha}
         dr, \; r\,d\theta, \; r\,\sin\theta\,d\phi \right\}
}
yields the \As\ potentials
\eqn\schasp{\eqalign{
        A_1 &= \left({M\over r^2} - {\lambda\over 3}r\right) d\tau + %%
                    \cos\theta\,d\phi \cr
     A_2 &= -\left(1 - {2M\over r} - {\lambda\over 3}r^2\right)^{\ha}
\sin\theta
\,d\phi \cr
   A_3 &= \left( 1 - {2M\over r} - {\lambda\over 3}r^2\right)^{\ha} d\theta \cr
}}
and the matrix $S$:
\eqn\schs{
    S = \hbox{diag}\,\left( -{2M\over r^3} - {\lambda\over 3}, \; {M\over r^3}
-
     {\lambda\over 3}, \; {M\over r^3} - {\lambda\over 3} \right)
}
This solution is therefore of Type D generally, for $M \neq 0$, and of
Type O in the limit of vanishing mass.   One may view the mass term as
a parameter which breaks the system out of the Type O sector.
When $\lambda \to 0$, we recover the usual Schwarzschild solution.

Reversing the orientation gives
\eqn\schbar{
    \overline{A}_1 = -\left({M\over r^2} - {\lambda\over 3}r\right)d\tau
              - \cos \theta d\phi \; = - A_1
}
with
$$
     \overline{A}_2 = A_2 \qquad \hbox{and} \qquad \overline{A}_3 = A_3
$$
while the form of $S$ is preserved.

The Pontrjagin index for $\lambda = 0$ can be computed to give
\eqn\schpont{\eqalign{
     P_1 &= -{1\over 2\pi^2}\int \tr \left(S^2 \right)\left(*1\right) \cr
         &= - {1\over \pi^2}\int^{2\pi}_{\phi = 0} \int^{\pi}_{0} %%
         \int^{\infty}_{r= 2M}\int^{8\pi M}_{\tau = 0} \; %%
               {3M^2\over r^4} d\tau\w dr \w \sin\theta d\theta \w d\phi \cr
       &= - 4 \cr
       &= \overline{P}_1 \cr
}}
The radius $2M$ is the usual event horizon, and we have also used
the periodicity in the Euclidean time interval of $8\pi M$ inherent
in the Schwarzschild metric.    From Eqn.\schpont,
we conclude that $\chi = 2$, and $\tau = 0$, in agreement with the
standard result.

Finally, note that a general Type D metric with zero cosomological constant
can be characterized by $S = \hbox{diag}(-2\alpha, \alpha, \alpha )$.  If
$\alpha > 0$, $S$ is gauge equivalent to
\eqn\schhiggs{
    S_{ab} = {1\over 3}\phi^2 \delta_{ab} - \phi_a \phi_b
}
since this form can be diagonalized to
$$
     \hbox{diag}\;\left( -{2\over 3}\phi^2 , \; {1\over 3}\phi^2 , \; %%
           {1\over 3} \phi^2 \right)
$$
For the Schwarzschild solution, $\phi ^2 = 3M/r^3$.   In isotropic
coordinates, with
$$
    r \equiv \left(1 + {M\over 2 \rho}\right)^2 \rho
$$
the quantity $\phi$ above takes on the value
$$
    \phi^a = \left(3M\right)^{\ha} \left(\rho\right)^{-{5\over 2}}%%
            \left(1 + {M\over 2\rho}\right)^{-3} \rho^a
$$
yielding for the \As\ magnetic field
\eqn\schmag{
     B^{ia} = {M\over \rho^3 \left(1 + {M\over 2\rho}\right)^{6}}\left(%%
            \delta^{ab} - {3\rho^a\rho^b \over \rho^2} \right)\ts^i _b
}
This establishes the gauge-equivalence between the Schwarzschild solution
in \As\ variables in our general formalism and the solution exhibited in
\ref\FKS{T.\ Fukuyama and K.\ Kamimura, Mod.\ Phys.\ Lett. {\bf A6}, 1437
(1991)
.}.

%\bye
\def\f{\left[1 - \left({a\over R}\right)^4 - {\lambda\over 6}R^2 \right]}
\def\ff{\left[1 + \left({a\over R}\right)^4 - {\lambda\over 12}R^2 \right]}
\bigskip
\line{%
(d)\underbar{The Eguchi-Hanson metric} \hfil}
\medskip
The Eguchi-Hanson metric
\ref\eh{T.\ Eguchi and A.\ Hanson, Phys. Lett. {\bf 74B}, 249
(1978);
Ann.\ Phys.\ {\bf 120}, 82(1979).} with a cosmological constant
can be written as
\eqn\ehmet{\eqalign{
  ds^2 &= {\f}^{-1}dR^2 + R^2\left(\Theta_1\,^{2} + \Theta_2\,^2 \right)  \cr
        &\quad + R^2{\f}\Theta_3\,^2 \cr
}}

We can choose the vierbein fields to be
\eqn\ehvier{
    e_A = \left\{ {\f}^{-\ha}dR,\, R\Theta_1 , \, R\Theta_2, \, %%
         {\f}^{\ha}R\Theta_3 \right\}
}
which then implies that the \As\ potentials are given by
\eqn\ehash{\eqalign{
     A_1 &= -2{\f}^{\ha} \Theta_1 \cr
     A_2 &= -2{\f}^{\ha} \Theta_2 \cr
     A_3 &= -2{\ff}^{\ha} \Theta_3 \cr
}}

The corresponding matrix $S_{ab}$ takes the form
\eqn\ehs{
    S = \hbox{diag}\,\left({4a^4\over R^6} - {\lambda\over 3},\, %%
   {4a^4\over R^6} - {\lambda\over 3}, \, -{8a^4\over R^6} - {\lambda\over 3}%%
              \right)
}
The Eguchi-Hanson metric is therefore of Type D when $a \neq 0$, and of
Type O when $a = 0$, so this parameter causes the system to break out of the
Type O sector.

When we apply a reversal, we get another manifold, $\overline{EH}$, with
the \As\ potentials taking the form:
$$
     \overline{A}_1 = \overline{A}_2 = 0
$$
while
\eqn\ehbarp{
     \overline{A}_3  = -{\lambda\over 2}R^2 \Theta_3
}
The field strengths are now controlled by the matrix
\eqn\ehbars{
     \overline{S} = \hbox{diag}\,\left( 0, \, 0, \, -\lambda\right)
}
Like in the case of $\overline{CP}_2$,  this matrix is not invertible,
and it is described by an abelian anti-instanton.

However, the Eguchi-Hanson manifold has a boundary of real projective 3-space,
$RP_3$\eh.  The abelian instanton of Eqn.\ehbarp\
does not depend on the parameter $a$, and furthermore, it is anti-self-dual
relative to $\overline{EH}$ for arbitrary $\lambda$ and $a$.    In the
limit $\lambda \to 0$, $S$ becomes zero, and $\overline{EH}$ becomes
half-flat.    As we shall see below, the Eguchi-Hanson metric can be
obtained as limiting cases of two different classes of explicit solutions,
one from the $F=0$ sector, and the other from the abelian anti-instanton
sector.

%\bye
\bigskip
\centerline{%
{\bf B. New Solutions}}
\bigskip
The above examples illustrate the procedure for determining the appropriate
anti-instanton configuration of the \As\ variables once the metric is
known.   But, the formalism can be used to go the other way and yield new
solutions to the Einstein field equations.   We shall illustrate the method
below by examining a few explicit examples.

Before we do so, recall that the matrix $S$ for Riemannian manifolds is
real-symmetric.   Solutions are characterized by $\Tr S^2$ and $\Tr S^3$,
which can be further divided into classes relative to a sign change under
orientation reversal.    This distinction had been utilized in the examples
presented so far, and will continue to be significant in the solutions
we will be discussing below.
\bigskip
\line{%
\underbar{{\bf $F = 0$ sector and hyperk\"ahler manifolds}}
\hfil}
\bigskip
We first examine the case where the \As\ field strength vanishes.  When this
happens, the metric is half-flat; {\it i.e.} the Riemann curvature is
self-dual.    $S$ vanishes also, and for simply-connected manifolds,
we may set the connection to be zero globally as well.   The equations
of motion reduce to
\eqn\sd{
           d\,\Sigma_a = 0
}
so that the anti-self-dual $\Sigma_a$ is now also closed.
As a result,
\eqn\btwo{
    b_2\,^{-} = 3
}
Since the \As\ curvature vanishes, we obtain
\eqn\pone{
    0 = P_1 = 3\tau (\cM) - 2 \chi (\cM)
}
so that $\tau$ takes on the maximal value of the Hitchin bound:
\eqn\sig{
    \tau (\cM) = {2\over 3}\chi (\cM)
}
But we also have the relation
$$
     \chi (\cM) = b_2 + 2 = b_2\,^{+} + b_2\,^{-} + 2 = b_2\,^{+} + 5
$$
so that finally
\eqn\sigf{
     \tau (\cM) \equiv b_2 ^{+} - b_2\,^{-} = b_2\,^{+} - 3
}
These relations may be solved to give the following characteristic numbers
for simply-connected compact Einstein manifolds in the $F=0$ sector:
\eqn\charn{
    b_2\,^{+} = 19 \qquad b_2\,^{-} = 3 \qquad \tau = 16 \qquad \chi = 24
}

It is known that $K3$ manifolds and the 4-torus are the only compact manifolds
without boundary admitting metrics of self-dual Riemann curvature\ref\EGH{T.
Eguchi, P. B. Gilkey and A. J. Hanson, Phys. Rep. {\bf 66},213 (1980).}.
The 4-torus is not simply-connected, and has $\tau = \chi = 0$, since
its metric is flat.  So, choosing the convention that $\tau(K3) = -16$,
we can identify the simply-connected compact half-flat manifolds without
boundary as $\overline{K3}$.    They have the intersection form\donkron:
\eqn\intf{
      {\cal Q} = \bigoplus^{3} \left[ \matrix{ 0 & 1 \cr  %%
                                     1 & 0 \cr } \right] %%
                 \bigoplus^{2}  E_8
}

The Pontrjagin index for $\overline{K3}$ can be computed to be
\eqn\pontb{
     \overline{P_1} = - 3\tau - 2 \chi = -96 \quad (=0 \; \hbox{\rm mod}4\;)
}
As a result, the $SO(3)$ \As\ connection can be lifted to an $SU(2)$
connection.
The metric therefore possesses an $SU(2)$ holonomy, and is therefore
hyperk\"ahler\SA.
Such metrics
have been used to formulate conditions for unbroken supersymmetry in the
compactification of superstrings\ref\GSW{M.\ B.\ Green, J.\ Schwarz and
E.\ Witten, {\it Superstring theory}, (Cambridge University Press, Cambridge,
1987).}.  In our present context, these metrics are associated
with the unbroken topological field theory of the moduli space of flat
connections\csbrst.

Note that although $F=0$ and $S=0$ for $K3$, the corresponding values for
$\overline{K3}$ need not be trivial.  It has been calculated that
these surfaces are parametrized by 58 parameters\ref\PA{D.\ N.\ Page,
Phys.\ Lett. {\bf 80B}, 55(1978).}.   According to Eqn.\ \pontb,
these must be associated with an \As\ connection with Pontrjagin number $-96$.

We shall now construct explicitly half-flat Einstein manifolds which
are not necessarily simply-connected, or without boundary.   They will
have $F=0$, but $\overline{F} \neq 0$.

We begin by supposing that the vierbein is of the form:
\eqn\viersd{
      e_A = \left\{ - a(R)dR, \, f(R)\Theta_1 ,\, g(R)\Theta_2 ,\, %%
                  h(R)\Theta_3 \right\}
}
This yields
\eqn\sdash{\eqalign{
     A_1  &= \left\{{f'\over a} - {(g^2 +  h^2 - f^2 )\over gh}\right\}%%
                     \Theta_1 \cr
    A_2 &= \left\{{g'\over a} - {(h^2 + f^2 - g^2 )\over fh} \right\}\Theta_2
\cr
    A_3 &= \left\{{h'\over a} - {(f^2 + g^2 - h^2 )\over fg}\right\}\Theta_3
\cr
}}
where primes denote differentiation with respect to $R$.  Further
simplification
can be achieved by assuming that $f=g$.  Setting $A_a = 0$ locally,
we need to solve
$$
        {f'\over a } = {h\over f}
$$
and
$$
      {h'\over a } + {h^2 \over f^2 } = 2
$$
Combining these two equations gives
$$
    {(h^2 )'\over (f^2 )'} + {h^2\over f^2 } = 2
$$
With $u\equiv h^2$ and $v\equiv f^2$, this equation reduces to
$$
     (uv)' = (v^2)'
$$
which has as solution
\eqn\sol{
   h^2 = f^2 + {b\over f^2}
}
with $b$ being an integration constant.   The metric is therefore given by
\eqn\metsd{
    ds^2 = a^2 dR^2 + f^2 \left(\Theta_1\,^2 + \Theta_2\,^2 \right) + %%
                    h^2 \Theta_3\,^2
}
where $a= (f^2)' /2h$, and $h$ is given Eqn.\ \sol.   The function
$f$ is an arbitrary function of $R$.

If we now reverse the orientation, the metric is invariant, but the
vierbein changes to $\overline{e_A} = (-e_0, e_i )$.  The \As\ potentials
become
\eqn\abar{\eqalign{
    \overline{A}_1 &= -{2h\over f} \Theta_1 \cr
    \overline{A}_2 &= -{2h\over f} \Theta_2 \cr
    \overline{A}_3 &= -2\left(2 - {h^2\over f^2 }\right) \Theta_3 \cr }
}
assuming that the relations among $h, a$ and $f$ continue to hold.
A short computation then fixes the matrix $S$ to be:
\eqn\smat{
    \overline{S} = \hbox{diag}\;\left(-{4b\over f\,^6 }, \; -{4b\over f\,^6},%%
                      \; {8b\over f\,^6} \right)
}
Thus the equations of motion still hold, but the solution is now of Type D
when $b\neq 0$.    For compact manifolds without boundaries,
\eqn\pontbar{\eqalign{
       \overline{\cal P}_1 &= -{1\over 4\pi^2}\int
\overline{F}_i\w\overline{F}_
i \cr
              \quad     &= 12b^2 \int (f\,^{-8})' dR \cr }
}
assuming that the variables $\theta, \phi$ and $\psi$ are the coordinates
of a 3-sphere for fixed values of $R$.   By choosing the appropriate
function $f$, one can obtain self-dual Einstein manifolds with non-trivial
values of the Pontrjagin number.   For the special case of $f=R$, and
$b= -a^4$, we recover the $\lambda \to 0$ limit of the Eguchi-Hanson
metric discussed above.   Recall that our convention is such that
$\overline{EH}$ with $\lambda = 0$ is half-flat.
\vfil\eject
\line{%
\underbar{{\bf Abelian anti-instantons and K\"ahler-Einstein manifolds}}
\hfil}
\bigskip
When $S = \hbox{diag}\,(0, \, 0, \, -\lambda )$, the \As\ potential is
described by an abelian anti-instanton.  In this gauge, the only non-vanishing
component of the field-strengths is $F_3$,   and the equations of motion
reduce to
\eqn\abeleq{\eqalign{
    dA_3 &\; = F_3 = -\lambda\, \Sigma_3 \cr
    d\Sigma_1 &\; = A_3 \w \Sigma_2 \cr
    d\Sigma_2 &\; = -A_3 \w \Sigma_1 \cr
}}
We now suppose that the manifold can support a complex structure, and
define
\eqn\compsig{
     \Sigma^{+} \equiv \Sigma_1 + i\,\Sigma_2
}
in which case,
\eqn\compeqn{
            d\Sigma^{+} + i\,A_3 \w \Sigma^{+} = 0
}
Furthermore, let us define
\eqn\compvier{\eqalign{
    \Omega^1 & \equiv -e^0 + i\, e^3 \cr
    \Omega^2 & \equiv e^1 + i\, e^2 \cr
}}
Then,
\eqn\compcl{
    \Sigma^3 = {i\over 2}\Omega^{\alpha}\w\overline{\Omega}^{\alpha} \qquad %%
                 \al = 1,2
}
is closed, by Eqn.\ \abeleq.
Therefore, Einstein manifolds which are endowed with a complex
structure, and are described by abelian \As\ anti-instantons can be
identified as K\"ahler manifolds, with $\Sigma_3$ as the K\"ahler form.

We now construct explicit solutions for this class of manifolds.  We
shall assume that the vierbein fields are of the form Eqn.\ \viersd,
and that the \As\ potentials satisfy Eqn.\ \sdash.    The equations
Eqn.\ \abeleq, it can be checked, are then satisfied.    We shall
now suppose, for simplicity, that
\eqn\simp{
     A_3 = c(R) \Theta_3
}
so that
\eqn\simpf{
   F_3 = c'\,dR \w \Theta_3 + 2\,c\,\Theta_1\w\Theta_2
}
To satisfy the gauge condition on $S$ as specified above, {\it i.e.}
diag$(0,\,0,\, -\lambda )$, we must have
\eqn\solns{
    c' = -\lambda a\, h   \qquad\quad 2c = -\lambda\,f\, g
}
It is easy to check that for $f=g$, Eqn.\ \solns\ implies that
both $A_1$ and $A_2$ vanish, and we are left with the condition
\eqn\cond{
   {h'\over a} - {(2f^2 - h^2)\over f^2 } = c
}
Substituting for $f^2$ from Eqn.\ \solns\ gives
$$
   -\lambda\left( h^2 c\right)' = {2\over 3}\left(c^3 \right)' + 2\left(%%
                      c^2 \right)'
$$

The solution is
\eqn\solnh{\eqalign{
      h^2 &= -{2\over 3\lambda }\left\{c(c+3) + {b\over c}\right\} \cr
      a^2 &= {(c')^2\over \lambda^2 h^2 } \cr
      f^2 &= g^2 = -{2c\over \lambda }  \cr
}}
The function $c$ is an arbitrary function of $R$, while $b$ is an
integration constant.

Upon reversal of orientation, the new \As\ variables are
\eqn\bara{\eqalign{
    \overline{A}_1 &= -{2f'\over a}\Theta_1 \cr
    \overline{A}_2 &= -{2f'\over a}\Theta_2 \cr
    \overline{A}_3 &= \left(-{c\over 3} - 2 + {2b\over 3\,c^2 }\right)\Theta_3
\cr
}}
The connections above now describe a non-abelian anti-instanton, with
the corresponding $S$ matrix given by
\eqn\bars{
    \overline{S} = \hbox{diag}\;\left( - {\lambda\over 3}\left[ 1 - {2b\over
c^3
}\right], %%
       \; -{\lambda\over 3}\left[ 1 - {2b\over c^3}\right], \, %%
        -{\lambda\over 3}\left[ 1 + {4b\over c^3 }\right] \right)
}
As a result, the solution is of Type D for $b\neq 0$, and of Type 0 when
$b=0$.

The corresponding Pontrjagin numbers are
\eqn\ptrj{\eqalign{
    {\cal P}_1  &= {1\over 4\pi^2} \int F_a \w F_a \cr
      \quad   &= - \int \left(c^2 \right)' dR \cr
}}
for the case of the abelian anti-instanton, and
\eqn\ptrjbar{\eqalign{
    \overline{\cal P}_1 &= {1\over 4\pi^2} \int \overline{F}_a \w
\overline{F}_a
 \cr
           \quad &= -\int \left({c^2\over 3 } + {8b\over 9c}
                 - {4b^2\over 3c^4}\right)' dR \cr
}}
for the non-abelian case.

If we let
$$
    c= - {\lambda R^2 \over 2\left(1 + {\lambda\over 6}R^2\right)}, \qquad
   b=0
$$
as an example, we reproduce the expressions for $CP_2$ and $\overline{CP}_2$.
Another example, with $b\neq 0$ is obtained by setting
$$
     c= -{\lambda\over 2}R^2,  \qquad b= -{3\over 4}\lambda^2 a^4
$$
This ansatz gives us the Eguchi-Hanson space and the configuration for
$\overline{EH}$ discussed in the last section.
\bigskip
\newsec{%
Matrix S as an order parameter}
\bigskip
We have seen how $S$ can play an effective role as an order parameter
characterizing the Type O sector.   This sector corresponds classically
to conformally self-dual Einstein manifolds.    Actually, we can go
further with this hypothesis by studying it in the abelian anti-instanton
sector. We have already discussed several explicit examples of Einstein
manifolds which belong to this sector.

Suppose that $S$ is of rank one and can be expressed as
\eqn\higgs{
     S_{ab} = \pm\phi_a\phi_b
}
where $\phi_a$ is a triplet of phenomenological
real scalar fields. It is then gauge-equivalent to the form
$S = \hbox{diag}( 0, 0, \pm \phi^2 )$ and we may assume that
\eqn\vev{
    ||\phi|| = \sqrt{\mp \lambda}
}
where the sign in Eqn.\ \higgs  is chosen in accordance with
whether $\lambda$ is negative or positive. In the $U$-gauge, with
$\phi_a = ||\phi||\delta_{a3}$, and $A_{1,2}=0$, we have simply the
condition
\eqn\covc{
    \left(D\phi\right)_a = 0
}
But Eqns.\ \vev and \covc\ are gauge and diffeomorphism invariant statements,
and are therefore valid in arbitrary $SO(3)$ gauges and coordinate systems.
The situation is therefore identical to that of a system possessing a
symmetry based on the group $SO(3)$, which is broken down to $SO(2)$ by the
order parameter $\phi_a$ acquiring a non-vanishing vacuum expectation value
equal to the cosmological constant. The matrix $S$ is non-invertible, and
in this phase the gravitational fields are ordered dynamically in such a
way as to break the local $SO(3)$ Ashtekar symmetry.

%\bye
%\input harvmac
\newsec{
Concluding Remarks
}
\bigskip
We have presented in this paper several examples which illustrate the
methods to be used in obtaining solutions to the Einstein equations
with Ashtekar variables.    The examples have been chosen to bring
out those features which are particularly transparent within this
context.   Among these are the properties of Einstein manifolds under
orientation reversals  and their relations to
abelian anti-instantons,  the role  of the cosmological
constant in fixing the the type of Einstein manifolds, and finally,
a perspective on spontaneous breaking of the local $SO(3)$
symmetry.   We hope to  amplify upon some of the physical
implications of these features, especially in a quantum context, in
the near future.

\vfil\eject
\centerline{%
{\bf Figure Caption}}
\bigskip
\item{Fig.\ 1} Classification of the initial data according to $S$.
\vfil\eject

\listrefs
\bye